%% file: main.tex
\documentclass{article}
\usepackage{geometry}
\usepackage{amsmath,amssymb,graphicx}

\def\x{{\mathbf x}}
\def\y{{\mathbf y}}
\def\a{{\mathbf a}}

\def\R{\mathbb{R}}
\def\h{{\mathbf h}}
\def\B{\text{B}_\text{ctc}(\y)}
\def\A{\text{B}_\text{nt}(\y)}
\def\sos{\left<\text{s}\right>}
\def\blank{\left<\text{b}\right>}
\def\pred{\mathbf{q}}
\def\V{\mathcal{V}}
\DeclareMathOperator*{\argmax}{arg\,max}

\usepackage{xcolor}

\definecolor{darkspringgreen}{rgb}{0.09, 0.7, 0.27}
\newcommand{\weiran}[1]{{\color{blue} Weiran: #1}}

\newcommand{\RPP}[1]{{\color{brown} RPP: #1}}

\renewcommand{\weiran}[1]{{}}
\renewcommand{\RPP}[1]{{}}

\title{Massive End-to-end Models for Short Search Queries}
\author{Weiran Wang \and Rohit Prabhavalkar \and Dongseong Hwang \and  Qiujia Li \and Khe Chai Sim \and Bo Li \and James Qin \and Xingyu Cai \and
Adam Stooke \and Zhong Meng \and CJ Zheng \and Yanzhang He \and Tara Sainath \and Pedro Moreno Mengibar \\
Google LLC \\
\texttt{\{weiranwang,prabhavalkar,dongseong\}}@google.com}

\begin{document}
%
\maketitle
\begin{abstract}
In this work, we investigate two popular end-to-end automatic speech recognition (ASR) models, namely Connectionist Temporal Classification (CTC) and RNN-Transducer (RNN-T), for offline recognition of voice search queries, with up to 2B model parameters. The encoders of our models use the neural architecture of Google's universal speech model (USM), with additional funnel pooling layers to significantly reduce the frame rate and speed up training and inference. We perform extensive studies on vocabulary size, time reduction strategy, and its generalization performance on long-form test sets.
Despite the speculation that, as the model size increases, CTC can be as good as RNN-T which builds label dependency into the prediction, we observe that a 900M RNN-T clearly outperforms a 1.8B CTC and is more tolerant to severe time reduction, although the WER gap can be largely removed by LM shallow fusion.
\end{abstract}

\input{intro}
\input{methods}
\input{architecture}
\input{experiments}
\input{conclusions}

\bibliographystyle{IEEEbib}
\bibliography{refs}

\end{document}

%% file: intro.tex
\section{Introduction}
\label{sec:intro}
There has been a recent focus on scaling up end-to-end ASR models, such as connectionist temporal classification (CTC)~\cite{graves2006connectionist} and neural transducers (RNN-T,~\cite{Graves_12a}), to extremely large sizes~\cite{zhang2022bigssl,li2022lifelong, radford2023robust, zhang2023google, pratap2023scaling} to explore the benefits of such models at scale.
Many of these models are motivated by the goal of training a single multilingual ASR model which can perform well across a range of languages.
For example, the Universal Speech Model (USM,~\cite{zhang2023google}) is a 2B parameter full-context CTC model trained on YouTube data from 300+ languages; Pratap et al.~\cite{pratap2023scaling} train ASR systems on 1000+ languages.
The present work follows in the footsteps of these previous works.
As large language models (LLMs) have rapidly risen in popularity in the field of natural language processing~\cite{openai2023gpt4, touvron2023llama, anil2023palm}, it is natural to ask how ASR models operate at large sizes.
Specifically, how do the two main model classes -- CTC and RNN-T -- compare at large model sizes? 
In particular, how do these models compare on a task (voice search, in this work) where paired training data is plentiful, for a single language (English) instead of in a multilingual setup.

A related, and important practical question relates to the difficulty of training massive end-to-end models as model size increases.
Since the cost of computing the outputs of the encoders in the model (most of the model parameters are typically in the encoder) scales 
poorly in attention-based encoders such as transformers~\cite{Vaswani_17a} and conformers~\cite{gulati2020conformer}, which can be prohibitive in terms of computation and memory.
The RNN-T model, in particular, requires additional computation and memory for the prediction and joint network, which further compounds the problem.

In this work, we investigate the aforementioned issues and find that the two issues -- how CTC and RNN-T compare, and how one can train them efficiently -- are not as independent as might seem initially: we find that encoder output frame rate reduction can be applied repeatedly at multiple layers in the encoder to obtain a large output frame rate reduction; this in turn is critical for training models efficiently.
While past research incorporated time reduction to lower the final frame rate to 60ms for voice search~\cite{he2019, ding2022}, we find that we can increase time reduction to 6x with large CTC, and all the way up to 16x with large RNN-T, of the 40ms base frame rate with funnel pooling~\cite{dai2020funnel}, without sacrificing much accuracy.


Additionally, we compare CTC and RNN-T as the backbone ASR system. CTC produces label probabilities that are independent of previous time frames, and is thus much faster at inference compared to RNN-T. This is one reason why CTC was preferred for USM in~\cite{zhang2023google}. However, we find that RNN-T is actually more accurate than CTC, and specifically a 900M RNN-T clearly outperforms a 1.8B CTC, even though the CTC model benefits significantly from increased model size compared with a 340M CTC. Moreover, we find that RNN-T can tolerate a much larger time reduction factor. In order to address any concerns with RNN-T being slower than CTC, we run both models (including beam search) on a TPU to achieve the best possible latency.

Finally, with the much smaller number of output frames and on-TPU beam search, shallow fusion~\cite{gulcehre2015using,Chorowski2017} with a neural LM becomes a viable candidate for enhancing model accuracy. We fuse both CTC and RNN-T with external LMs trained with large amounts of text, and observe that the word error rate (WER) gap between them can be largely removed by LM fusion. Our study on shallow fusion is timely, as the community starts to become interested in the use of LLMs for ASR. We show that shallow fusion remains an effective technique for LM integration, even for large models trained with hundreds of thousands of hours of audio, and should be considered as a baseline for more advanced techniques.


%% file: methods.tex
\section{End-to-end ASR Models}
\label{sec:methods}
\vspace*{-1ex}

In this section, we briefly describe the end-to-end models employed in this work -- connectionist temporal classification (CTC)~\cite{graves2006connectionist}, and neural transducers~\cite{Graves_12a} -- particularly,  the hybrid autoregressive transducer (HAT)~\cite{variani2020hybrid}.
Since end-to-end models are now part of the mainstream in automatic speech recognition (ASR), we opt for brevity; interested readers can find more information in recent overview articles
~\cite{li2022recent, prabhavalkar2023end}.

\noindent\textbf{Notations} We assume that the input audio signal has been parameterized into suitable acoustic feature vectors: $\x = [\x_1, \ldots, \x_{T'}]$, of length $T'$, where $\x_t \in \R^{d}$ (128-dimensional log-mel features, in this work). 
The input acoustic features are processed using an \emph{encoder}, a suitable neural network (a Conformer~\cite{gulati2020conformer}, in this work), which transforms the input into a higher-level representation: $\h = [\h_1, \ldots, \h_T]$, where the length of the encoded representation is typically shorter than the original input length ($T \leq T'$).
We assume that we have an input transcript corresponding to each utterance: $\y=[y_0=\sos, y_1, \ldots, y_U]$, where each $y_u \in \V$, the set of output symbols (word-pieces~\cite{wordpieces}, in this work), and $\sos$ represents a special start-of-sentence symbol. 

\subsection{Connectionist Temporal Classification}
\vspace*{-1ex}

Connectionist Temporal Classification (CTC) was introduced by Graves et al.~\cite{graves2006connectionist}, as a way to train sequence-to-sequence models which can transduce the input sequence $\x$, into the output sequence $\y$ when the alignments between the two sequences are unknown.
CTC accomplishes this by modeling the conditional distribution, $P(\y|\x)$, by marginalizing over all possible alignment paths between the two sequences:
\begin{equation}
    P(\y|\x) = P(\y | \h(\x)) = \sum_{\a \in \B} \prod_{t=1}^{T} P(a_t |\h) \label{eq:ctc}
\end{equation}
where, the $\B$ corresponds to the set of all valid alignments.
Specifically, an alignment $\a \in \B$ is a valid alignment if it contains $|\h(\x)| = T$ symbols from the set of outputs augmented with a special \emph{blank} symbol -- $\V \cup \{\blank\}$; and if additionally, removing consecutive repeated non-blank symbols and then removing all $\blank$ symbols produces the original label sequence $\y$.
As can be seen in~\eqref{eq:ctc}, CTC models make a strong conditional independence assumption -- that the output labels are conditionally independent given the input acoustic encoded features. 
However, these models work well in practice~\cite{miao2015eesen, Karita2019}, especially with large encoders~\cite{zhang2023google}.
Finally, as can be seen in~\eqref{eq:ctc}, the CTC model produces one output symbol (blank, or non-blank) per encoder time step.

\subsection{Neural Transducers}
\vspace*{-1ex}

The recurrent neural network transducer (RNN-T) was proposed by Graves et al.~\cite{Graves_12a}, as an improvement over the CTC model, to reduce conditional independence assumptions in the model.
This is achieved by introducing a separate \emph{prediction network}, which models label dependency (and thus makes model outputs conditionally dependent on the sequence of previous predictions). 
\begin{equation*}
    P(\y|\x) = P(\y | \h(\x)) = \sum_{\a \in \A} \prod_{\tau=1}^{T+U} P(a_\tau | \pred_{i_\tau}, \h_{\tau - i_\tau})
\end{equation*}
where, $\A$ is the set of all valid alignments -- sequences of length $T + U$, which are identical to $\y$ after removing all blanks (i.e., each sequence has exactly T blank symbols); $i_\tau$ represents the number of non-blank labels in the partial alignment $a_1, \ldots, a_{\tau-1}$; and, $\pred_{j}$ represents the output of the prediction network after processing the first $j-1$ output labels: $\pred_j = \text{PredNetwork}(y_{j-1}, \ldots, y_0)$.
While early works employed recurrent LSTM networks to model the prediction network, in this work, the prediction network is modeled as $|V|^2$ embedding network~\cite{Rami21}, whose output only depends on the last two non-blank labels $y_{j-1}$, and $y_{j-2}$.
Notably, neural transducers allow the model to output multiple (non-blank) output labels at each frame; blanks correspond to transitions to the next input encoder frame.

\subsubsection{Hybrid Autoregressive Transducer}
\vspace*{-1ex}

The hybrid autoregressive transducer (HAT) model proposed by Variani et al.~\cite{variani2020hybrid} improves over the basic neural transducer structure in two ways. 
First, it factorizes the output distribution $P(a_\tau | \pred_{i_\tau}, \h_{\tau-i_\tau})$ into two separate distributions: a Bernoulli distribution to model blank symbols (i.e., if $a_\tau = \blank$), and a separate probability distribution for the non-blank labels.
Second, the model proposes the notion of an internal language model (ILM), $P_\text{ILM}(\y)$, which represents the LM learned by the model based on the training data; the ILM can thus be subtracted out from the posterior distribution, before fusion with an external LM, $P_\text{EXT}(\y)$, during decoding, by adding tunable hyperparameters $\alpha$, and $\beta$ which can be set based on a development set to produce the most likely hypothesis~\cite{morgan90,variani15,kanda17,mcdermott19}:
\begin{equation*}
\y^{*} = \argmax_\y \log P(\y|\x) - \alpha \log P_\text{ILM}(\y) + \beta \log P_\text{EXT}(\y)
\end{equation*}

%% file: architecture.tex
\section{Model architecture}
\label{sec:architecture}
\subsection{USM Conformer}
\vspace*{-1ex}

The USM uses the convolution-augmented Transformer~\cite{gulati2020conformer}, or Conformer architecture. Each Conformer block consists of a feed-forward module ($\textsc{FFN}$), a multi-head self-attention module ($\textsc{MHSA}$), a convolution module ($\textsc{Conv}$), and a second feed-forward module. If the input to $l$-th Conformer block is $\mathbf{x}^{(l)}$, then its output (and, the input to the next block), $\mathbf{x}^{(l+1)}$, is computed as:
\begin{align}
\nonumber
    \tilde{\mathbf{x}}^{(l)} & = \mathbf{x}^{(l)} + \frac{1}{2}\textsc{FFN}_{1}\big(\mathbf{x}^{(l)}\big)
            \end{align}
    \begin{align}\label{eq:conformer}
    \mathbf{x}'^{(l)} & = \tilde{\mathbf{x}}^{(l)} + \textsc{MHSA}\big(W_{q}\tilde{\mathbf{x}}^{(l)},
    W_{k}\tilde{\mathbf{x}}^{(l)}, W_{v}\tilde{\mathbf{x}}^{(l)}\big) \\
    \nonumber
    \mathbf{x}''^{(l)} & = \mathbf{x}'^{(l)} + \textsc{Conv} \big(\mathbf{x}'^{(l)}\big) \\
    \nonumber
    \mathbf{x}^{(l+1)} &= \textsc{LayerNorm}\big(\mathbf{x}''^{(l)} + \frac{1}{2}\textsc{FFN}_{2}(\mathbf{x}''^{(l)})\big)
\end{align}
For the standard conformer layer, the input length and output length are the same.

\subsection{Funnel Pooling}
\vspace*{-1ex}

We improve training and inference speed in our models by reducing the encoder embedding sequence length relative to the audio length.  At selected Conformer blocks within the encoder, we employ the pooling technique in the \textsc{MHSA} module introduced in Funnel-Transformers \cite{dai2020funnel}.  At a funnel self-attention layer, the entire input is used to produce the key and value sequences as usual, but strided-average pooling is applied to the time dimension of the input used in producing the query vectors.  
That is, we replace the MHSA module in~\eqref{eq:conformer} by \eqref{eq:pooling} below.
\begin{equation}
    \label{eq:pooling}
    \begin{split}
        \hat{\mathbf{x}}^{(l)} & = \textsc{StridedPooling}(\tilde{\mathbf{x}}^{(l)}), \\
        \mathbf{x}'^{(l)} & = \hat{\mathbf{x}}^{(l)} + \textsc{MHSA}\big(W_{q}\hat{\mathbf{x}}^{(l)}, W_{k}\tilde{\mathbf{x}}^{(l)}, W_{v}\tilde{\mathbf{x}}^{(l)}\big).
    \end{split}
\end{equation}
The stride of pooling in~\eqref{eq:pooling} is the time reduction factor in this layer.
In a language model, this pool-query-only method was shown to provide a slight advantage over simply pooling the entire hidden embedding sequence between layers~\cite{dai2020funnel}; likely, the higher granularity in the key and value sequences allows the network to learn a less lossy compression.  Computational savings accrue in all subsequent Conformer layers according to the $O(T^2D)$ complexity for self-attention, and linearly for feed-forward networks.

\noindent\textbf{Related work on time reduction}
Time reduction has been a useful technique for achieving a balance between input and output lengths for ASR, and has evolved over time with the choice of neural architecture and modeling units. In~\cite{vanhoucke2013,miao2016}, the lower frame rates were achieved by concatenating input frames and striding, for DNN or LSTM models predicting context-dependent HMM states.
\cite{hihi1996, koutnik14, Chan_16a} proposed to use hierarchical/pyramidal RNNs where outputs of consecutive steps are combined before feeding to the next layer. 
After the community switched to end-to-end systems and word-piece type modeling units, a popular frame rate is 40ms as achieved by convolutional subsampling, and is adopted by widely-used open-source libraries~\cite{watanabe2018espnet}. With the Conformer architecture~\cite{gulati2020conformer}, previous work on voice search mostly had a final frame rate of 60ms, achieved by a 30ms input frame rate and a 2x time reduction layer early in the encoder, with stacking~\cite{he2019} or funnel pooling~\cite{ding2022}.
With regards to the loss function, recent work of~\cite{wang2023ctcreduction} proposed to use CTC to (irregularly) select encoder output frames for RNN-T modeling, effectively reducing the frame rate for the decoder. However, such a CTC-based encoder output selection does not save any computations in the encoder, but only reduces the computation in downstream modules.
Our goal in this work is to aggressively reduce the sequence length early in the architecture for computational savings.

%% file: experiments.tex
\section{Experiments}
\label{sec:experiments}

\subsection{Datasets}
\label{sec:data}
\vspace*{-1ex}

Our experiments focus on short voice search queries. For a majority of the experiments, we use 520M utterances of voice search queries for training; the total amount of audio is 490K hours and the average duration per utterance is 3.4 seconds. A small percentage of the training data is human transcribed while the rest is pseudo-labeled by a 600M bidirectional RNN-T teacher~\cite{hwang2022}. We tokenize training transcripts with word-piece models~\cite{wordpieces}.

We use both real audio and TTS-generated data for evaluation. The real audio utterances are representative of typical voice search traffic, with an average duration of 3.9 seconds. Our development set consists of 9K real audio utterances (denoted as VS-dev); we use a separate held-out test set consisting of 5K utterances for testing (denoted as VS-test).
The TTS sets contain rare proper nouns (RPN) which appear fewer than 5 times in the training set, and they are good testbeds for external LM integration.
Each TTS set contains 10K utterances and covers one of five domains: Maps (denoted as RPNM), News (RPNN), Play (RPNP), Search query logs (RPNS), and Youtube (RPNY); they have average durations of 5.9, 10.1, 5.3, 5.4, and 5.8 seconds respectively. We use VS-dev, RPNM and RPNN for tuning the model architecture and other hyperparameters and report final WERs on the rest sets.

For training LMs for fusion, each minibatch is sampled 50/50 from the transcripts of acoustic training data, and text-only data which contains 50B utterances. The text-only data contains textual search queries from the domains of Maps, Textual search query logs, News, Play, and Youtube, and a frequency-based pruning strategy, designed to improve rare word modeling, is implemented to adjust the probability of selecting each query~\cite{huang2022sentence}. We train transformer LMs of 128M and 1B parameters (not counting parameters in the final softmax) with wordpieces compatible with those of E2E models.

\subsection{Model Architectures}
\label{sec:expt_architecture}
\vspace*{-1ex}

We use the 128-dimensional log Mel-filterbank energies (extracted from 32ms window and 10ms shift) as the frontend features. After two 2D-convolution layers, both with strides (2,2), the resulting feature sequence has a frame rate of 40ms and becomes the input to our Conformer architecture. This architecture mimics that of Google's universal speech model (USM,~\cite{zhang2023google}). The number of attention heads used in Conformer blocks is 8, and the intermediate dimension of the FFNs is 4 times the model dimension. The Conformer blocks use local self-attention with a large attention span, and the encoder output has a large enough receptive field to cover the entire utterance. As observed by~\cite{zhang2022bigssl}, unsupervised pre-training does not help with ASR accuracy when using a large amount of supervised audio data, and thus we skip the pre-training step in this work (except for including a baseline CTC model with pretraining in Table~\ref{tab:ctc}).

We explore two different encoder configurations for CTC, with different sizes: the smaller encoder configuration consists of 24 Conformer layers of dimension 768, leading to a total of 340M parameters; the larger encoder has 32 Conformer layers of dimension 1536, leading to a total of 1.8B parameters.
For RNN-T, the encoder consists of 16 Conformer layers of dimension 1536, resulting in a total of 870M parameters. We use a $|V|^2$ embedding decoder~\cite{Rami21}, i.e., the prediction network computes LM features based on two previous non-blank tokens, which was shown to work well on voice search data. The model output uses the HAT factorization~\cite{variani2020hybrid} which was shown to benefit external LM integration.

Deviating from the USM, we perform significant time reduction in the encoder architecture. Following~\cite{ding2022}, we initially start funnel pooling at layer index 4 (zero-based), and apply pooling in subsequent layers to achieve the desired factor of time reduction. As an example, if we perform pooling at layers with (zero-based) indices 4 and 5, each with the reduction factor 2, we achieve a total reduction factor 4 for layers 6 and onwards, i.e., the sequence length after layer 6 is 1/4 of the original input length and the frame rate at the encoder output is 160ms, which is also the frame rate at which the decoder operates.
Note that~\cite{ding2022} used funnel pooling for on-device modeling with stringent latency requirements, whereas here we use funnel pooling for the full-context model to speed up training and inference. We observe that the ASR accuracy turns out to be more tolerant of time reduction. In Sec~\ref{sec:start_pooling}, we investigate the effect of starting funnel pooling at earlier layers to further reduce inference costs.

Each model is trained with the Adafactor optimizer~\cite{adafactor18} with a batch size of 4096 utterances. We train CTC models to 500K steps and RNN-T models to 300K steps, by which point their WERs on development sets have stabilized. In a pilot study, we have conducted discriminative training with the minimum word error rate objective~\cite{prabhavalkar2018}. We achieved no WER gain on VS sets and 3\% relative improvement on the RPN sets. We only present results with maximum likelihood training in this work, and it is future work to carefully study discriminative training for large models.

For CTC, we perform greedy decoding when not using external LM, in which case beam search does not provide additional WER gain. We do perform prefix beam search~\cite{hannun2014} with a beam size of 8 when fusing with external LM. Two types of prior probabilities were used for CTC shallow fusion: the uniform prior over non-blanks as implemented by the blank probability downscaling technique~\cite{sak2015}, versus the uni-gram prior based on the model posteriors on training set~\cite{li2019}.
We perform label synchronous beam search (with path merging) for RNN-T with a beam size of 8, and perform internal LM score subtraction in the case of shallow fusion.

\begin{table}[t]
    \centering
    \caption{WERs (\%) on dev sets by CTC, with different architectures and vocabulary sizes. Funnel pooling is employed at layers 4 and 5, with factors 2,2 for 160ms frame rate and 3,2 for a 240 frame rate.}
    \vspace{1em}
    \label{tab:ctc}
    \begin{tabular}{|c|c|c|c|} 
    \hline
    Model size (frame rate) & VS-dev & RPNM & RPNN \\ \hline \hline
    \multicolumn{4}{|c|}{baselines: no pooling, vocab size=4096} \\ \hline
    340M (40ms) & 4.7 & 15.1 & 12.4 \\
    1.8B (40ms), w. pretraining & 4.2 & 14.2 & \bf 10.4 \\
    \hline \hline
    \multicolumn{4}{|c|}{with funnel pooling, vocab size=16384} \\ \hline
    340M (160ms)  &  4.5 & 15.3 & 16.6 \\ 
    340M (240ms)  &  4.5 & 14.9 & 21.6  \\
    \hline
    1.8B (160ms)  &  4.3 & \bf 13.7 & 12.7 \\ 
    1.8B (240ms)  &  \bf 4.2 & 13.8 & 17.3 \\
    1.8B (320ms)  &  5.0 & 14.0 & 26.4 \\
    \hline
    \multicolumn{4}{|c|}{1.8B (160ms) + 128M LM fusion} \\ \hline
    blank downscaling~\cite{sak2015} & 3.8 & 10.7 & 11.0 \\
    model-based prior~\cite{li2019} & 3.8 & 10.5 & 10.8 \\ \hline
    \multicolumn{4}{|c|}{1.8B (160ms) + 1B LM fusion} \\ \hline
    blank downscaling~\cite{sak2015} & 3.8 & 10.1 & 10.4 \\
    model-based prior~\cite{li2019} & \bf 3.8 & \bf 9.8 & \bf 10.1 \\
    \hline
    \end{tabular}
    \vspace{-1em}
\end{table}



\subsection{CTC Results}
\label{sec:ctc_results}
\vspace*{-1ex}

Intuitively, with a larger vocabulary, the label sequences are shorter and we can afford more time reduction. The hard constraint for CTC is that, since it emits only one token (blank or non-blank) at each encoder output frame, the encoder output sequence length must remain longer than the label sequence length (RNN-T however is not subject to this constraint). During CTC training, we discard utterances that violate this constraint, although such cases are uncommon in the VS sets. For example, for the 16K vocabulary size, at the final frame rate of 320ms (8x reduction), roughly 1\% utterances violate this constraint as estimated on VS-dev.

We report WERs of a selection of CTC models in Table~\ref{tab:ctc}.
As baselines, we report the WERs with 4K vocabulary at a 40ms frame rate, a setup closely following that of the USM architecture~\cite{zhang2023google}. We match these baselines on VS with the 16K vocabulary size and much lower frame rates of 160ms and 240ms: for 340M CTC, we obtain an improvement from 4.7\% to 4.5\% on VS-dev, while for 1.8B CTC, we match the 4.2\% without pretraining. For supervised training with 1.8B CTC, we observe a 4x speedup in training time with a 160ms frame rate, and a 4.5x speedup with a 240ms frame rate compared to the 40ms model (despite the use of a more costly softmax operation due to larger vocabulary). However, we do observe that with a 320ms frame rate, the WER on VS-dev significantly degrades to 5.0\%, and this could not be alleviated by increasing the vocabulary size to 32K. This suggests that a too coarse time resolution does not work well with the CTC loss and its underlying independence assumptions.

\begin{table}[t]
    \centering
    \caption{Dev set WERs (\%) by RNN-T with various vocabulary sizes and frame rates.  The encoder contains 870M parameters. Funnel pooling starts at layer 4 and applies to adjacent layers. We use the notation 3x2 to indicate that layer 4 has a reduction factor of 3, and layer 5 has a reduction factor of 2, which yield the final frame rate of 240ms.}
    \vspace{1em}
    \label{tab:rnnt}
        \begin{tabular}{|l|c|c|c|} 
    \hline
    Frame rate (reduction factors) & VS-dev & RPNM & RPNN \\ \hline \hline
    \hline
    \multicolumn{4}{|c|}{vocab size=4096,\ \ \ decoder size=10M} \\ \hline
    240ms (3x2)   & 3.8 & 12.6 & \bf 12.1 \\
    320ms (2x2x2) & 3.8 & 12.9 & 14.2 \\
    \hline \hline
    \multicolumn{4}{|c|}{vocab size=16384,\ \ \ decoder size=33M} \\ \hline
    160ms (2x2)  &  \bf 3.7 & \bf 12.3 & 12.8 \\ 
    240ms (3x2) & 3.8 & 12.5 & 12.2 \\
    320ms (2x2x2) & 3.8 & 12.5 & 13.6 \\
    400ms (5x2) & 3.8 & 12.7 & 15.6 \\
    480ms (3x2x2) & 3.9 & 12.6 & 19.5 \\
    640ms (2x2x2x2) & 3.9 & 12.9 & 28.7 \\ \hline \hline
    \multicolumn{4}{|c|}{vocab size=32768,\ \ \ decoder size=65M} \\ \hline
    320ms (2x2x2) & 3.8 & 12.9 & 14.8 \\
    640ms (2x2x2x2) & 3.9 & 12.7 & 25.9 \\
    \hline \hline
    \multicolumn{4}{|c|}{vocab size=16384,\ \ \ 128M LM fusion}  \\ \hline
    160ms (2x2) & 3.8 & \bf 11.0 & 32.6  \\ \hline
    \end{tabular}
    \vspace{-1em}
\end{table}

The trend of WER on RPNM is similar to that of VS-dev. However, we do observe worse degradation on RPNN as we apply heavier time reduction. As mentioned in Sec~\ref{sec:data}, the transcriptions of RPNN come from the News domain which has different linguistic characteristics from voice search, and the audio length is quite longer (10 secs on average) than the VS set (4 secs on average). We hypothesize that models with lower frame rates may not generalize well to unseen audio length, and further investigate this issue in Sec~\ref{sec:long_form}.

Given that the recent work on large models is based on CTC~\cite{radford2023robust,zhang2023google}, one may speculate that as the bidirectional CTC model gets larger, and with large amounts of training data, the underlying modeling assumption of CTC holds approximately and it can achieve state-of-the-art accuracy by itself. We challenge this speculation by performing LM shallow fusion to the 1.8B CTC model with a 160ms frame rate. We observe a significant WER reduction from 4.2\% to 3.7 -- 3.8\% with a smaller 128M LM already on the in-domain VS-dev set. The results we present in Table~\ref{tab:ctc} uses LM weights that achieve a good balance between VS-dev and RPN sets. Had we focused only on VS, the best we could achieve on VS-dev is 3.7\%, with clearly worse WERs on RPN sets. Among the two prior estimation methods, model-based prior~\cite{li2019} consistently outperforms blank probability downscaling~\cite{sak2015} over different WER operating points. When increasing the external LM size to 1B, we could not further improve on VS-dev but achieved sizeable gains on rare word sets. 

\RPP{I'm not very convinced by this paragraph. The LM has context, so the joint decoding function doesn't make independence assumptions -- i.e., CTC alone makes conditional independence assumptions, but CTC + LM is not conditionally independent. If you think this is an important point to make, we can, but I think we need to be more careful.}
\weiran{What I wanted to say is that the independent assumption is not very good. Applying the LM fusion which is not conditionally indepenent improves the accuracy.}

\subsection{RNN-T Results}
\label{sec:rnnt_results}
\vspace*{-1ex}

We conduct a similar set of time reduction experiments with RNN-T and the results are shown in Table~\ref{tab:rnnt}. Overall RNN-T is quite robust to different vocabulary sizes.
With the 16K vocabulary, we studied the most time reduction settings, and we observed only small WER degradations from 160ms frame rate all the way to 640ms frame rate on VS-dev and RPNM. Like CTC, the performance on the ``out-of-domain'' dataset RPNN degrades as the frame rate reduces. Comparing the models at the same frame rate of 160ms and vocabulary size 16K, RNN-T has a WER of 3.7\% on VS-dev, outperforming the 4.2\% by CTC by a large margin, and achieving parity with CTC + shallow fusion on this set. This demonstrates the benefit of having a learnable LM feature encoder for modeling label dependency in end-to-end ASR.

When fusing the best RNN-T model with a 128M LM, we observe interestingly that it tends to significantly degrade RPNN (with heavy deletion errors) and the WERs are quite sensitive to internal and external LM weights. We list one set of results in Table~\ref{tab:rnnt} (bottom panel) which achieves a good balance between VS-dev and RPNM, yet it degrades RPNN from 12.8\% to 32.6\%. We hypothesize this is due to the bias of the internal language model of HAT learned purely on short utterances. In Sec~\ref{sec:long_form}, we provide further evidence for this,  by demonstrating that training on length-diverse data improves the WER and robustness to shallow fusion on RPNN. 
\RPP{We should add something like, .... In Section X.X, we provide further evidence for this fact by demonstrating that training on length-diverse data improves ...
}\weiran{added a explanation.}

\begin{table}[t]
    \centering
    \caption{Dev set WERs (\%) by 900M RNN-T with a vocab size of 16K and a 240ms frame rate (3x2 reduction), with funnel pooling started in earlier Conformer layers. The first row is taken from Table~\ref{tab:rnnt}.}
    \vspace{1em}
    \label{tab:funnel_start_layer}
        \begin{tabular}{|c|c|c|c|} 
    \hline
    Start layer index & VS-dev & RPNM & RPNN \\ \hline \hline
    4 & 3.8 & 12.5 & 12.2 \\
    3 & 3.7 & 12.6 & 12.0 \\
    2 & 3.7 & \bf 12.4 & \bf 11.7 \\
    1 & 3.7 & 12.6 & 12.3  \\
    \hline
    \end{tabular}
    \vspace{-1em}
\end{table}

\subsection{Location of Funnel Pooling}
\label{sec:start_pooling}
\vspace*{-1ex}

So far, we have started funnel pooling from layer 4, following prior work~\cite{ding2022} that worked on smaller models with hard latency constraints. In this section, we start pooling earlier in the architecture, which leads to more efficient inference. 

We use the 240ms frame rate RNN-T model from Table~\ref{tab:rnnt} as the baseline, with time reduction factors of 3 and 2 in two consecutive layers, and change the pooling start layer index to 3, 2, and 1.
\RPP{Is there a more effective way to describe this? I find the funnel pooling descriptions throughout this paper to be quite confusing.}
Training of the model with pooling starting at layer 0 diverged with the same learning parameters and we do not report its performance. The results of these models on development sets are reported in Table~\ref{tab:funnel_start_layer}, which shows that on top of the 40ms base frame rate, the model is quite robust to the location of pooling layers, and in fact we obtain small WER gains on RPN sets by starting pooling at layer 2. We plan to replace the 2D convolutional subsampling layers before the Conformer layer with funnel pooling layers to explore more possibilities for frame reduction.

\begin{table}[t]
    \centering
    \caption{Dev set WERs (\%) by 900M RNN-T with 16K vocabulary and two frame rates 240ms and 640ms trained on multi-domain data. Funnel pooling starts from layer 4. First two rows are taken from Table~\ref{tab:rnnt}.}
    \vspace{1em}
    \label{tab:long-form}
        \begin{tabular}{|c|c|c|c|} 
    \hline
    Frame rate & VS-dev & RPNM & RPNN \\ \hline \hline
    \multicolumn{4}{|c|}{Voice search training} \\ \hline
    240ms & 3.8 & 12.5 & 12.2 \\
    640ms & 3.9 & 12.9 & 28.7 \\ \hline \hline
    \multicolumn{4}{|c|}{Multi-domain training} \\ \hline
    240ms & \bf 3.6 & 13.6 & \bf 6.9 \\
    640ms & 3.8 & 12.6 & 9.5 \\ \hline
    \multicolumn{4}{|c|}{+ 128M LM fusion} \\ \hline
    240ms & 3.7 & \bf 11.5 & 7.1 \\ \hline \hline
    \multicolumn{4}{|c|}{Multi-domain 600M teacher, vocab size=4096} \\ \hline
    60ms & 4.0 & \bf 11.5 & 7.1 \\ 
    \hline
    \end{tabular}
    \vspace{-1em}
\end{table}

\subsection{Adding Long-Form Training Data}
\label{sec:long_form}
\vspace*{-1ex}

To verify that the poor performance of end-to-end models on RPNN was due to the lack of longer training audio, we repeat several RNN-T experiments with additional multi-domain training data~\cite{narayanan2019recognizing}. Most notably, we include segmented YT audio data containing 520M utterances with an average duration of 9.8 seconds, giving us a total of 600K hours of longer-form training data.

The comparisons between RNN-T models trained on voice search data and multi-domain data are presented in Table~\ref{tab:long-form}. For two vastly different frame rates 240ms and 640ms, the additional long-form training data improves the WERs on both VS-dev and RPNN by a lot. With only voice search training data, the WER gap on RPNN between the two frame rates used to be very large (12.2\% vs 28.7\%), whereas with multi-domain training, the gap is significantly reduced (6.9\% vs 9.5\%) with much improved absolute WERs.
When performing shallow fusion for the multi-domain 240ms frame rate model with the same 128M LM used in Sec~\ref{sec:rnnt_results}, we achieve a better balance between in-domain and out-of-domain test sets,  improving RPNM from 13.6\% to 11.5\% and without affecting RPNN much.
\RPP{I don't understand the footnote completely. Isn't the LM also trained on RPNN data? You mention that we sample text from all of the domains.}.
\weiran{I removed the footnote to be safer. You are right, text-only data should include RPNN, though there may not bee too much PRNN text-only data IIUC. So far we do not have long-form speech-text data for LM training which may improve the scenario.}

It is interesting to add the bidirectional RNN-T teacher~\cite{hwang2022} into comparison, as shown in the bottom panel of Table~\ref{tab:long-form}. The 600M teacher is trained on an earlier version of multi-domain data, where the voice search portion is smaller but fully hand-transcribed, and is used to generate the pseudo-labels of voice search data used in this paper. Note that we are surpassing or on par with the teacher's performance on VS and RPNN, probably due to a larger amount of training data and more consistent labeling. The teacher does outperform new models on RPNM, probably because, as an end-to-end model, it predicts even fewer rare words in its pseudo-labels than human transcriptions, for the student to imitate.

\begin{table}[t]
    \centering
    \caption{Test set WERs (\%) by CTC and RNN-T, with 16K vocabulary. Funnel pooling starts from layer 4. By default, we use VS training data.}
    \vspace{1em}
    \label{tab:test}
        \begin{tabular}{|c|c|c|c|} 
    \hline
    VS-test & RPNP & RPNS & RPNY \\ \hline \hline
    \multicolumn{4}{|c|}{1.8B CTC, 160ms frame rate} \\ \hline
    4.9 & 39.8 & 23.1 & 26.0 \\ \hline
    \multicolumn{4}{|c|}{+ 128M LM fusion with model prior} \\ \hline
    4.5 & 34.1 & 17.0 & 20.8 \\ \hline \hline
    \multicolumn{4}{|c|}{900M RNN-T, 240ms frame rate} \\ \hline
    4.5 & 37.8 & 20.6 & 23.3 \\ \hline
    \multicolumn{4}{|c|}{+ Multi-domain training data} \\ \hline
    4.4 & 36.4 & 19.9 & 22.3 \\ \hline
    \multicolumn{4}{|c|}{\ \ + 128M LM fusion with ILM} \\ \hline
    \bf 4.4 & \bf 33.9 & \bf 16.9 & \bf 20.2 \\
    \hline
    \hline
    \multicolumn{4}{|c|}{Multi-domain 120M RNN-T~\cite{ding2022}}\\
    \multicolumn{4}{|c|}{60ms frame rate, 0.9s look-ahead} \\ \hline
    5.0 & 35.9 & 19.2 & 23.2 \\
    \hline
    \end{tabular}
    \vspace{-1em}
\end{table}

\subsection{Final Evaluation}
\label{sec:test}
\vspace*{-1ex}

Finally, we compare the best configurations from both CTC and RNN-T on the test sets, namely VS-test, RPNP, RPNS, and RPNY. Taking into account both in-domain and out-of-domain performance, we choose the 1.8B CTC model with 160ms frame rate, and the 900M RNN-T model with 240ms frame rate. The results are shown in Table~\ref{tab:test}. Relative merits between methods are consistent with the performance on development sets. That is, with the same voice search training data, the 900M RNN-T outperforms the 1.8B CTC on all test sets, and by 8.0\% relative margin for the in-domain VS-test set (4.5\% vs 4.9\%).
For more stable shallow fusion performance on out-of-domain test sets, it is beneficial to train RNN-T on multi-domain data, and we expect the same for CTC. As a reference, we provide results of another small RNN-T model, with a 60ms frame rate and a limited right context of 0.9s, from previous work~\cite{ding2022}.

\RPP{What happens to a CTC model when trained on MD data? It feels like we need to include that number too.}
\weiran{I realized this but it was too late to train this model. I will do this expt in case reviewers ask.}

%% file: conclusions.tex
\section{Conclusions}
\label{sec:conclusions}
\vspace*{-1ex}

We have compared two major end-to-end ASR models, CTC and RNN-T, on a large-scale voice search task. We have verified that RNN-T is clearly more accurate than CTC for in-domain test data even with a smaller model size, although the gap can be largely removed with LM fusion which compensates for the label independence assumption underlying CTC.
We have also observed that, for large models, time reduction is effective at reducing inference and training costs without sacrificing accuracy, and is thus useful for offline transcription tasks.
In the future, we will further optimize the model architecture and extend the usage to much longer audio.